\documentclass[aps,showpacs,twocolumn,prl]{revtex4}

\usepackage{amsmath}
\usepackage{latexsym}
\usepackage{graphicx}

\begin{document}

\title{Gravitational wave extraction from an inspiraling configuration of merging black holes}

\author{John G. Baker, Joan Centrella, Dae-Il Choi, Michael Koppitz, James van Meter}
\affiliation{Gravitational Astrophysics Laboratory, NASA Goddard
Space Flight Center, 8800 Greenbelt Rd., Greenbelt, MD 20771, USA}

\date{\today}

\begin{abstract}
We present new techniqes for evolving binary black hole 
systems which allow the accurate determination of gravitational waveforms 
directly from the wave zone region of the numerical simulations.  Rather than 
excising the black hole interiors, our approach 
follows the ``puncture'' treatment of black holes, but
utilzing a new gauge condition which allows the black holes to move 
successfully through the computational domain.
We apply these techniques to an inspiraling binary,
modeling the radiation generated 
during the final plunge and ringdown.  We demonstrate convergence
of the waveforms and good conservation of mass-energy,
with just over 3\% 
of the system's mass converted to gravitional radiation.

\end{abstract}

\pacs{
04.25.Dm, 
04.30.Db, 
04.70.Bw, 
95.30.Sf, 
97.60.Lf  
}

\maketitle

Coalescing comparable mass black hole binaries are prodigious sources of
gravitational waves. The final merger of these systems, in which
the black holes leave their quasicircular orbits and plunge together
to produce a highly distorted black hole that ``rings down'' to a
quiescent Kerr state, will produce a strong burst of gravitational
radiation.  Such mergers are expected to be 
among the strongest sources
for ground-based gravitational wave detectors, which will observe 
the mergers
of stellar-mass and intermediate mass black hole binaries, and
the space-based
LISA, which will detect mergers of massive black hole binaries.
Observations of these systems will provide an unprecedented look
into the strong-field dynamical regime of general relativity. 
With the first-generation of ground-based interferometers reaching
maturity and LISA moving forward through the formulation phase, 
the need for accurate merger waveforms has become urgent. 

Such waveforms
can only be obtained through 3-D numerical relativity simulations 
of the full Einstein equations.  
While this has proven to be a very challenging undertaking,
new developments allow an optimistic outlook.  Full 3-D 
evolutions of binary black holes, in which regions within
the horizons have been excised from the computational grid,
have recently been carried out.  Using co-rotating coordinates,
so that the holes remain fixed on the grid as the system
evolves, a binary has been evolved through a little more
than a full orbit \cite{Bruegmann:2003aw} as well as through a plunge,
merger, and ringdown \cite{Alcubierre:2004hr}, though without being
able to extract gravitational waveforms.
More recently, a simulation in which excised black holes move
through the grid in a single plunge-orbit, merger, and ringdown
has been accomplished, with the calculation of a 
waveform \cite{Pretorius:2005gq}.

In this {\em Letter}, we report the results of new simulations of
inspiraling binary black holes through merger and ringdown.  These
have been carried out using new techniques which allow the black
holes to move through the coordinate
grid without the need for excision {\footnote{While this paper was
being written, we learned that M. Campanelli and collaborators had
independently developed similar techniques for moving black holes
without excision, when both groups presented their
results at the Numerical Relativity 2005 workshop 
(http://astrogravs.nasa.gov/conf/numrel2005/presentations/)
on Nov. 2, 2005.  Since then, Campanelli et al. submitted a manuscript
describing their work to the physics archive \cite{Campanelli:2005dd}.}}.
 Using fixed mesh refinement, we
are able to resolve both the dynamical region where the black holes
inspiral (with length scales $\sim M$, where $M$ is the total
system mass) and the outer regions
where the gravitational waves propagate (length scales $\sim (10 - 100)M$).
Using an outer boundary at $128M$, we evolve the system to well
beyond $t \approx 100M$,
extract gravitational waveforms and demonstrate that they are
$2^{\rm nd}$-order convergent.

We start by setting up ``puncture'' initial data for equal
mass binary black holes \cite{Brandt97b}.  The metric on the initial spacelike
slice takes the form $g_{ij} = \psi^4\delta_{ij}$, where $i,j = 
1,2,3$, and the conformal factor $\psi = \psi_{\rm BL} + u$.
The static, singular part of the conformal factor has the form
$\psi_{\rm BL} = 1 +  \sum_{n=1}^{2} m_n/2 |\vec{r} - \vec{r}_n|$,
where the $n^{\rm th}$  black hole has mass $m_n$ and
its located at $\vec{r}_n$.  The nonsingular function $u$ is
obtained by solving the Hamiltonian constraint equation
using {\tt AMRMG} \cite{Brown:2004ma}.
We use parameters so that the black holes
have proper separation $4.99M$, and the system has total mass
$M = 1.008$ and angular momentum $J = 0.779M^2$.  This corresponds
to the run QC0 studied in Ref.~\cite{Baker:2002qf}.

In the standard puncture implementation, 
$\psi_{\rm BL}$ is factored out and handled analytically; only the regular
parts of the metric are evolved. In this case, 
the punctures remain fixed on the grid while the binary evolves.
However, the stretching of the
coordinate system that ensues is problematical.
First, as the physical distance between the black holes shrinks,
certain components of the metric must approach zero,
causing other quantities to grow uncontrollably.  Second, a
corotating coordinate frame (implemented by an appropriate angular 
shift vector) 
is necessary to keep the orbiting punctures
fixed on the grid; this causes extremely superluminal coordinate
speeds at large distances from the
black holes
and, in the case of a Cartesian grid, incoming noise from the outer boundary.

Our approach is to allow the punctures to move freely through the grid, by
not factoring out the singular part of the conformal factor but rather evolving
it inseparably from the regular part. 
Initially, we follow
the standard puncture technique and set up the binary
so that the centers of the black holes are not located at a grid point.
Taking numerical derivatives of $\psi_{\rm BL}$
effectively regularizes the puncture singularity
using the smoothing inherent in the finite differences. These 
regularized data are then evolved numerically. 
Since the centers of the black holes remain in the $z=0$ plane, they
do not pass through gridpoints in our cell-centered implementation.

We evolve this data with the Hahndol code, which uses
a conformal formulation of Einstein's
evolution equations on a cell-centered numerical grid \cite{Imbiriba:2004tp}
with a box-in-box resolution
structure implemented via Paramesh \cite{MacNeice00}
The innermost refinement region is a cube stretching from $-2M$ to
$2M$ in all 3 dimensions, and has
the finest resolution $h_f$. The punctures are placed within this region
on the $y-$axis in the $z=0$ plane; we impose equatorial symmetry throughout.
We performed three simulations
with identical grid structures, but with uniformly differing resolutions.
In the most refined cubical region the resolutions 
were $h_f=M/16$, $M/24$, and $M/32$.
Subsequent boxes of doubled size
have half the resolution. We use $8$ boxes to put the outer boundary at $128
M$, causally disconnected from the wave extraction region through most of the
run. 
We use $4$th order finite differencing for the
spatial derivatives except for the advection of the shift, which is
performed
with $2^{\rm nd}$-order, mesh-adapted differencing~\cite{Baker:2005xe}, and we use $2^{\rm nd}$-order time stepping via a three-step iterative
Crank-Nicholson scheme.

In our new approach, the free evolution of punctures is made possible by 
a modified version of a common coordinate condition known as
the Gamma-freezing shift vector, which drives the coordinates 
towards quiescence as 
the merged remnant black hole also becomes physically quiescent.  
Our modified version
retains this ``freezing'' property, yet is suitable for motile
punctures.
 Specifically we use
$\partial_t \beta^i = \frac{3}{4} \alpha B^i$ and
$\partial_t B^i = \partial_t \tilde{\Gamma}^i -\beta^j
\partial_j\tilde\Gamma^i - \eta B^i$, which incorporates
two critical changes to the standard Gamma-freezing
condition.  A factor $\psi_{\rm BL}$
of the conformal factor, originally used to ensure that the shift
vanishes at the puncture, has been removed in order to allow the
punctures to move.  Also, a new term has been
added ($-\beta^j \partial_j\tilde\Gamma^i$) which facilitates more stable and accurate evolution of moving punctures by eliminating a zero-speed mode 
(which was otherwise found to create a ``puncture memory" effect as errors grew in place~\cite{BakerMoving}). 
Along with this shift condition, we use the
standard singularity-avoiding, 1+log slicing condition on the lapse.

\begin{figure}[t]
  \includegraphics*[width=18pc,height=14pc]{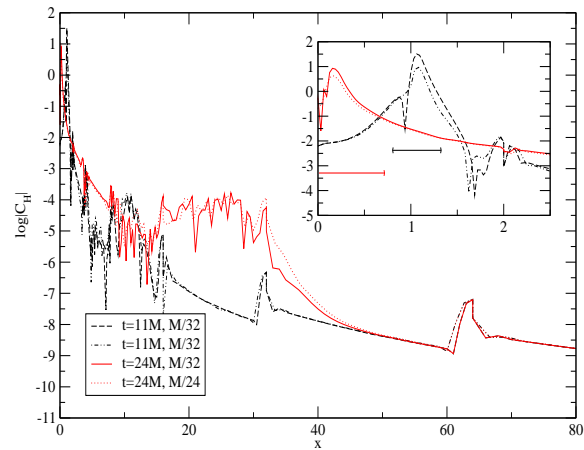}
  \caption{Hamiltonian constraint error $C_H$ for 
  $h_f=M/24$ and $M/32$, 
  at two
  times when a puncture is near to crossing the positive $x$-axis. 
  The
  data are scaled such that the lines should
  superpose in the case of perfect $2^{\rm nd}$-order convergence.
  The inset shows that $C_H$ is well-behaved in the region near the
  punctures. The horizontal lines indicate
 the approximate location of the apparent 
  horizons; at the later time a 
  common horizon has formed.
  \label{fig:HamConverge}}
\end{figure}

Fig.~\ref{fig:HamConverge} shows the
 error in the Hamiltonian constraint $C_H$ at 2 different times. 
The peak violation near the
puntures does not leak-out, or grow with time, but stays well-confined
even though the punctures and horizons move across the grid. 
Overall, we get $2^{\rm nd}-$order convergence away from the horizons 
to well beyond the wave extraction region for the entire course of the
run. There is no
indication of exponentially growing constraint violations which have
plagued many numerical simulations with black holes fixed in place on the
numerical grid.

\begin{figure}[t]
  \includegraphics*[trim=80 55 65 55,width=16pc,height=14pc]{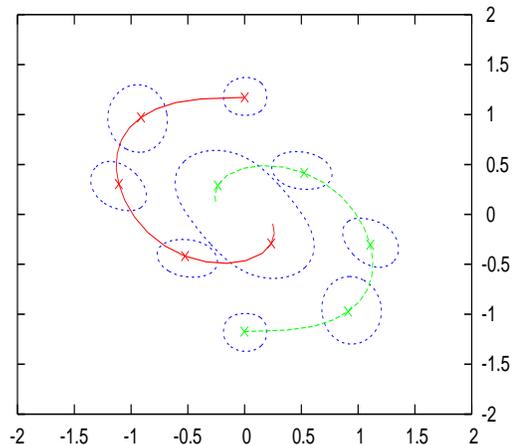} 
  \caption{The positions of the apparent horizons at times
$t=0,5,10,15$, and $20M$ for our $M/16$ run.  The curve shows the
trajectories of centroids of the individual apparent horizons.  }
  \label{fig:Horizons}
\end{figure}

One way to get a picture of the motion of the black 
holes is to look at the location of the black hole apparent horizons at 
different times. Fig.~\ref{fig:Horizons} shows the locations of a sequence
of apparent horizons (calculated using the 
{\tt AHFINDERDIRECT} code\cite{Thornburg:2003sf})
 where they cross the $x$-$y$-plane for our $h_f=M/16$ run.
 In the coordinates of our simulation, the
black holes undergo about one-half orbit before forming a common horizon.

We extract
the gravitational waves generated by the merger using the 
technique explained in detail in~\cite{Fiske:2005fx}. 
\begin{figure}[t]
  \includegraphics*[width=18pc,height=14pc]{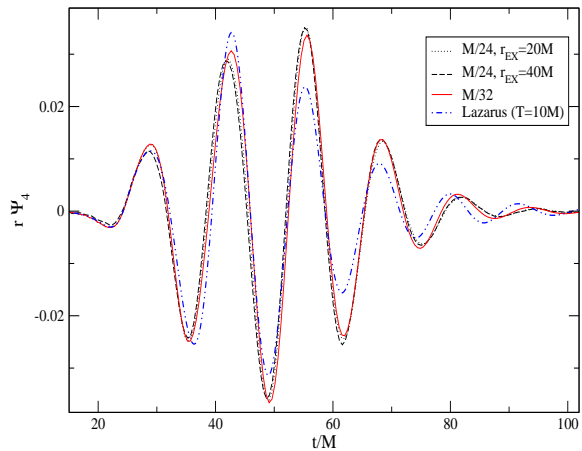} 
  \caption{Real part of $r \Psi_4$ extracted from the numerical
  simulation on spheres of radii $r_{EX}=20$, and $40M$ for the medium
  and high resolution runs.  The
  waveforms extracted at different radii have been rescaled by 
  $1/r_{EX}$ and
  shifted in time to account for the wave propagation time between the
  extraction spheres.  At high resolution ($h_f = M/32$) there is no discernible 
  dependence on extraction radius. For comparison, we show Lazarus 
  waveforms from  Ref.~\cite{Baker:2002qf}.}
  \label{fig:Waves}
\end{figure}
Fig.~\ref{fig:Waves} shows the dominant $l=2, m=2$ components of the
Weyl curvature scalar $\Psi_4$
extracted at 2 different radii from the medium and high resolution runs. 
For each resolution, the time-shifted and rescaled
waveforms computed at different extraction radii are nearly indistinguisable,
indicating that the waves travel cleanly
 across refinement boundaries and have the expected $1/r$ falloff.
In addition, the two highest resolution waveforms 
differ only by a slight phase shift, and a by few percent in 
amplitude.  For comparison we have also included the QC0 Lazarus
waveforms from Sec.~V of Ref.~\cite{Baker:2002qf}. These
were extracted by
approximately matching the later portion of brief numerical 
simulations onto a perturbed black hole \cite{Baker:2001nu,Baker:2001sf}
at transition time $10M$.

\begin{figure}[t]
  \includegraphics*[width=18pc,height=14pc]{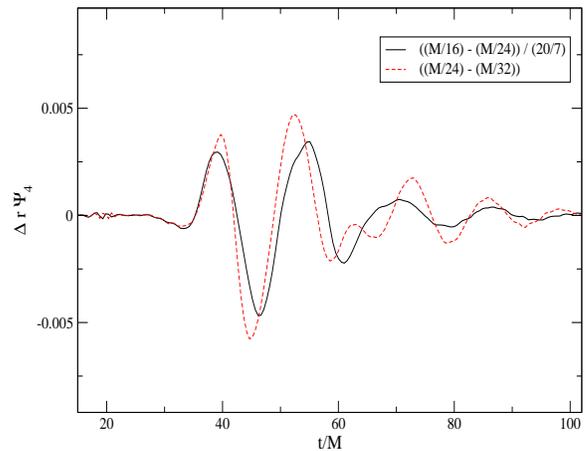} 
  \caption{Differences of the real part of $r \Psi_4$ for resolutions of
  $h_f=M/16, M/24$, and $M/32$ appropriately scaled
  such that for perfect $2^{\rm nd}$-order convergence the lines would lay on top
  of each other. }
  \label{fig:WaveConverge}
\end{figure}

Fig.~\ref{fig:WaveConverge} shows the convergence of the extracted waves
throughout the run. The difference between the two highest
resolutions is roughly $90$ degrees out of phase with the waveform,
corresonding to a small phase-shift in the waveform, possibly caused
by a small difference in the orbital trajectories.

The gravitational waveforms also contain 
physical information about the radiation, including the energy $E$ and 
angular momentum $J$ carried away by the radiation.  We calculate $dE/dt$ 
and $dJ/dt$ from time integrals of all $l=2$ waveform components using 
Eqs.~(5.1) and (5.2) in \cite{Baker:2002qf}.
Integrating $dE/dt$ gives the energy loss as a function of time; this should
be the same as $M - M_{\rm ADM}$, where $M_{\rm ADM}$ is the ADM mass
extracted at on a sphere of sufficiently large radius~\cite{Arnowitt62}.  
\begin{figure}[t]
  \includegraphics*[width=18pc,height=14pc]{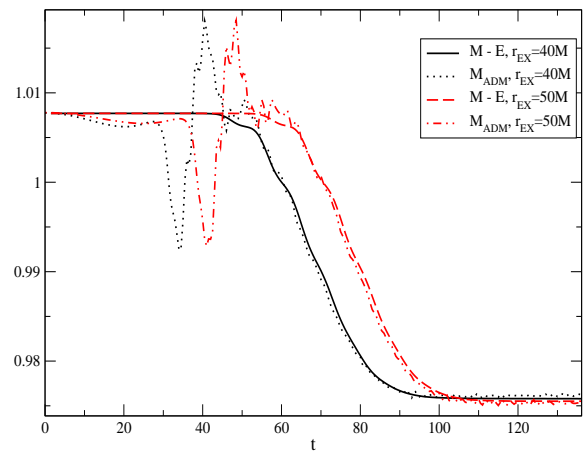} 
  \caption{Conservation of mass-energy for the highest resolution case,
    $h_f = M/32$.  We compare the ADM mass $M_{\rm ADM}$ with the 
    mass remaining, $M - E$, after gravitational radiation energy loss $E$.
    The good agreement, based on extraction spheres at $r_{EX}=40$ and $50M$,
    indicates conservation of energy in the simulation.} 
  \label{fig:EnergyCons}
\end{figure}
Fig.~\ref{fig:EnergyCons} shows a comparison between $M - E$ and
an independent calculation of $M_{\rm ADM}$ at two extraction 
radii $r_{EX}=40$ and $50M$.  
The striking consistency between the two calculations as the radiation passes
indicates good energy conservation in the simulation.  Shortly before the 
arrival of the radiation the ADM mass measurement is affected by a transient
non-physical pulse in the gauge evolution, though the pulse does not affect 
the radiation measurement.

The total radiated energy calculated 
from the waveforms extracted at $r_{EX}=20, 30, 40$ and $50M$ in the
highest resolution run has the values
$E/M=$0.0304, 0.0312, 0.0317 and 0.0319, respectively. 
While these values vary significantly with $r_{EX}$ (even extracting at 
these relatively large radii), they are neatly consistent 
with a $1/r_{EX}$ falloff to an asymptotic value 
of $0.0330$ with an 
uncertainty in the extrapolation of $< 1\%$. 
In Table~\ref{table:numbers}
we give the total radiated energies and angular 
momenta extrapolated as $r_{EX}\rightarrow\infty$.
\begin{table}[h]
  \begin{center}
  \begin{tabular}{ c c c c c c c}
  \hline
  \hline
   & $M/16$ & $M/24$ & $M/32$ & Lazarus & AEI\\ 
  \hline  
  $E/M$  $\;\;$ & 0.0516 $\;\;$   &  0.0342 $\;\;$ &  0.0330$\;\;$  & 0.025 $\;\;$   & 0.030\\
  $J/M^2$ & 0.208 $\;\;$    &  0.140 $\;\;$  &  0.138 $\;\;$  & 0.10  $\;\;$   & 0.17\\  
  \hline
  \hline
  \end{tabular}
  \end{center}
  \caption{The radiated energy E and angular momentum J carried away
  by gravitational radiation in our simulations.  Our values are
  comparable with earlier estimates from the AEI (via horizon
  analysis) and Lazarus {via perturbation techniques}.
  \label{table:numbers} }
\end{table}
For comparison we also include the Lazarus values, as well as values from
the AEI group~\cite{Alcubierre:2004hr} which did not 
determine waveforms, but estimated the
radiative losses based on the state of the final black hole horizon in runs 
including the QC0 case.  Our lowest resolution run clearly over-estimates the 
radiation energy and angular momentum while our higher resolution results 
are in closer agreement with the AEI value for the energy, and close to
 the 20\% level of 
confidence suggested in Ref.~\cite{Baker:2002qf}. 

In conclusion, we have calculated a gravitational radition waveform
directly via numerical simulation of an inspiraling configuration of
binary black holes.  Starting with equal mass puncture black holes and
parameters corresponding to the QC0 model of Ref.~\cite{Baker:2002qf},
we release the holes and allow them to move through the grid without
excision.  A new gauge condition is used that allows us to accurately
evolve this system from the initial inspiral orbit through merger and
ringdown.  The simulations converge with increasing resolution to
$2^{\rm nd}$-order, leading to a $2^{\rm nd}$-order convergence of the
waveform. These waveforms have the correct $1/r$ falloff and agree to
a great extent with approximatively calculated ones. Our simulations
show good energy conservation as indicated by comparing the change in
ADM mass with the radiated energy.  The QC0 configuration provides a
model for the final plunge of the two black holes and the subsequent 
ringdown.  In
this brief burst of gravitational radiation we find that just over 3\%
of the system's initial mass-energy is carried away in gravitational
waves.

The new gauge allows simulations to remain accurate far longer than previous
standard puncture
techniques. Our treatment will generalize, allowing us to study radiation 
generation in simulations of a variety of initial black hole configurations.
Using adaptive mesh refinement,  
we plan to apply these techniques to study binaries beginning from 
larger initial separation, which are expected to provide
more realistic models corresponding to astrophysical systems.  For further 
understanding of such model dependence, we will compare results from simulations
beginning with different initial data models.
We will also study the effects of unequal black hole masses, and the 
individual black hole spins.

\acknowledgments

We thank David Brown for providing {\tt AMRMG}; and
Jonathan Thornburg for providing {\tt AHFINDERDIRECT}, which 
we utilized with the assistance of Peter Diener and Thomas Radke. 
This work was supported in part by NASA grant ATP02-0043-0056.  The 
simulations were carried out using Project Columbia at NASA
Ames Research Center and at the NASA Center for
Computational Sciences at Goddard Space Flight Center.
M.K and J.v.M. were supported by the Research Associateship Programs
Office of the National Research Council.

\bibliographystyle{../bibtex/apsrev}

\bibliography{../bibtex/references}

\end{document}